\documentclass[5p,twocolumn,english]{elsarticle}
\usepackage{babel,amsmath,amssymb,dcolumn}
\usepackage{hyperref}

\usepackage{amsfonts}
\usepackage{amssymb}
\usepackage{graphicx}
\usepackage{latexsym}

\usepackage{dcolumn}

\newcommand{\be}{\begin{equation}}
\newcommand{\ee}{\end{equation}}

\begin{document}

\title{Signature of the interaction between dark energy and 
dark matter in galaxy clusters}

\author[ifusp]{Elcio Abdalla}
\ead{Email:eabdalla@fma.if.usp.br }

\author[ifusp]{L. Raul Abramo}
\ead{Email:abramo@fma.if.usp.br }

\author[iagusp]{Laerte Sodr\'e Jr.}
\ead{Email:laerte@astro.iag.usp.br }

\author[fu]{Bin Wang}
\ead{Email:wangb@fudan.edu.cn } 

\address[ifusp]{Instituto de F\'{\i}sica, Universidade de S\~ao Paulo,
  CP 66318,  
05315-970, S\~ao Paulo, Brazil} 

\address[iagusp]{Departamento de
Astronomia, Instituto de Astronomia, 
Geof\'isica e Ci\^encias Atmosf\'ericas da Universidade de S\~ao Paulo \\
Rua do Mat\~ao 1226, Cidade Universit\'aria, 05508-090, S\~ao Paulo,
Brazil }

\address[fu]{Department of Physics, Fudan University, 200433
  Shanghai, China}

\begin{abstract}
We investigate the influence of an interaction between dark
energy and dark matter upon the dynamics of galaxy clusters.
We obtain the general Layser-Irvine equation in the presence of
interactions, and find how, in that case, the virial theorem 
stands corrected. 
Using optical, X-ray and weak lensing data from 33 relaxed galaxy
clusters, we put constraints on the strength of the coupling between 
the dark sectors.  Available data suggests that this coupling is small but 
positive, indicating that dark energy might be
decaying into dark matter. Systematic effects between the several
mass estimates, however, should be better known, before definitive
conclusions on the magnitude and significance of this coupling could be
established.
\end{abstract}


\maketitle

\section{Introduction}

Cosmological accelerated expansion is by now a well-established
observational fact \cite{cosmoaccel,wmap,bao}, leading either to
an asymptotically de Sitter cosmology, plagued with an
astonishingly small cosmological constant, or else to a
universe filled up to 80\% with a strange dynamical component
with negative pressure -- dark energy \cite{darkenergy}.

If dark energy contributes a significant fraction of
the content of the Universe, it is natural, in the framework
of field theory, to consider its interactions with
the remaining fields of the Standard Model and well-motivated
extensions thereof. For lack of
evidence to the contrary, interactions of dark energy or
dark matter with baryonic matter and radiation must be 
either inexistent or negligible.
Nevertheless, some level of interaction between dark energy
and the dark matter sector, which is present in most
extensions of the Standard Model, is still allowed by
observations.

The possibility that dark energy and dark matter can
interact has been studied in \cite{amendola}-\cite{[14]},
among others. It has been shown that the coupling between
a dark energy (or �quintessence�) field and the
dark matter can provide a mechanism to alleviate the
coincidence problem \cite{amendola,[11]}. A suitable choice of the coupling,
motivated by holographic arguments, can also lead
to the crossing of the �phantom barrier� which separates
models with equations of state $w = p/\rho > -1$ from models
with $w < -1$ \cite{[12]} -- see also \cite{[14],[15]}. In addition, it
has been argued that an appropriate interaction between
dark energy and dark matter can influence the 
perturbation dynamics and affect the lowest multipoles of the
CMB spectrum, accounting for the observed suppression
of the quadrupole \cite{[13],[16]}.  The
strength of the coupling could be as large as the fine
structure constant \cite{[13],[17]}. Recently, it was shown
that such an interaction could be inferred from the expansion
history of the universe, as manifested in, e.g.,
the supernova data together with CMB and large-scale
structure\cite{[18]}. Nevertheless, the observational
limits on the strength of such an interaction remain weak
\cite{[19]}.

A complementary and fundamentally different way in which the coupling
between dark energy and dark matter can be checked
against the observations is through its impact on large-scale
structure. If dark energy is not a cosmological constant,
it must fluctuate in space and in time -- and,
in particular, if dark energy couples to dark matter, then
it must surely be dynamical. If that is the case, dark
energy affects not only the expansion rate, but the
process of structure formation as well, through density
fluctuations, both in the linear \cite{[11]}, \cite{[20]}-\cite{[23]} and
the non-linear \cite{[24],[25]} regimes. The growth of dark
matter perturbations can in fact be enhanced due to the
coupling between these two components \cite{[13],[14],[26]}.

Recently, it was suggested that the dynamical equilibrium
of collapsed structures would be affected by the
coupling of dark energy to dark matter, in a way that
could be observed in the galaxy cluster Abell A586 \cite{[27]}.
The basic idea is that the virial theorem is distorted by
the non-conservation of mass caused by the coupling.

In this paper we show precisely how the Layser-Irvine
equation, which describes the flow to virialization \cite{[28]},
is changed by the presence of the coupling, in such a
way that the final state of equilibrium violates the usual
virial condition, $2K + U= 0$, where $K$ and $U$ are respectively the
kinetic and the potential energies of the matter constituents
in an isolated system. We show that this violation
leads to a systematic bias in the estimation of
masses of clusters if the usual virial conditions are employed. Although
it is still possible that systematic errors from observations smear the
results, the fact that some shift in the mean value of the coupling for
two independent sets of observations (compared to the third set) may
signalize some new physics.

Even though the uncertainties associated with any individual
galaxy cluster are very large, by comparing the
naive virial masses of a large sample of clusters with their
masses estimated by X-ray and by weak lensing data, we
may be able to constrain such a bias and to impose 
tighter limits on the strength of the coupling than has
been achieved before.

\section{PHENOMENOLOGY OF COUPLED DARK
ENERGY AND DARK MATTER MODELS}

Quite generically, at the level of the cosmological background
an interaction between dark matter and dark energy
manifests itself as a source term in the continuity
equations of both fluids:
\begin{eqnarray}
\dot\rho_{dm} + 3 H \rho_{dm} &=& \psi 
\\ \nonumber
\dot\rho_{de} + 3 H \rho_{de} (1+w_{de}) &=& - \psi \; ,
\end{eqnarray}
where a dot denotes time derivative, $H$ is the expansion rate,
$\rho_{dm}$ and $\rho_{de}$ are respectively the energy densities of dark
matter and dark energy, and $w_{dm}$ and $w_{de}$ are their 
equation of state parameter. Notice that the
continuity equation still holds for the total energy density
$\rho_{Tot} = \rho_{dm} + \rho_{de}$.

Phenomenologically, one can describe the interaction
between the two fluids as an exchange of energy at a rate
proportional to the total energy density \cite{[8],[12]}:
\be
\label{psi}
\psi = \zeta H \rho_{Tot} \; .
\ee
We are interested in collapsed structures -- places where
the local, inhomogeneous density $\sigma$ is far from the 
average, homogeneous density $\rho$. In
that case the continuity equation for dark matter reads:
\be
\label{cont}
\dot\sigma_{dm} + 3H\sigma_{dm} + 
\vec\nabla \left( \sigma_{dm} \vec{v}_{dm} \right) = 
\zeta H \left( \sigma_{dm} + \sigma_{de} \right) \; ,
\ee
where $\vec{v}_{dm}$ is the peculiar velocity of dark matter particles.

In this work we will consider the local density of dark
energy to be proportional to the local density of dark
matter, $\sigma_{de} = b_{em} \sigma_{dm}$. 
If for a given model
the dark energy component is very homogeneous, 
$b_{em} \approx 0$. 
We do not consider the case where $b_{em}$ depends on
the size and mass of the collapsed structure -- although
this should probably 
happen in realistic models of structure formation with dark
energy \cite{[25]}. Hence, the continuity equation with dark
matter coupled to dark energy reads:
\be
\label{cont_eq}
\dot\sigma_{dm} + 3H\sigma_{dm} + 
\vec\nabla \left( \sigma_{dm} \vec{v}_{dm} \right) = 
\bar\zeta H \sigma_{dm} \; ,
\ee
where $\bar\zeta = \zeta (1+b_{em})$ is the effective coupling
in a virialized structure. Notice that different dark energy models
predict different levels of dark energy perturbations \cite{[25],[26]},
hence any constraints we derive from observations of
collapsed structures will be in some sense degenerate with
the perturbative properties of the dark energy sector.

\section{LAYZER-IRVINE EQUATION IN THE
PRESENCE OF COUPLING}

We will use Newtonian mechanics to derive equilibrium
conditions for a collapsed structure in an expanding Universe.
The acceleration due to the gravitational force is
given by:
\be
\label{accel}
\left( a \, \vec{v}_{dm} \right) \dot{} \, = \, - a \, \vec\nabla \varphi \; ,
\ee
where $a$ is the scale factor and $\varphi$ is the (Newtonian) gravitational
potential. Multiplying both sides of this equation
by $\sigma_{dm} a \vec{v}_{dm}$, integrating over the volume and using
the continuity Eq. (\ref{cont_eq}), we get that the left-hand side
becomes:
\be
\label{lhs}
\left( a^2 K_{dm} \right) \dot{} - a^2 \bar\zeta H K_{dm} \; ,
\ee
where the kinetic energy of dark matter is given by:
\be
\label{K}
K_{dm} = \frac12 \int \vec{v \,}^2_{dm} \sigma_{dm} dV \; .
\ee
The right-hand side of the equation, on the other hand,
becomes:
\be
\label{rhs}
(1+b_{em}) \left[ - a^2 \left( \dot{U}_{dm} + H U_{dm} \right) + 2
  \bar\zeta H a^2 U_{dm} \right] \; ,
\ee
where we have used the Poisson equation, the fact that
$\sigma_{Tot} = (1 + b_{em}) \sigma_{dm}$, and the definition
of the potential energy of a distribution of dark matter particles:
\be
\label{U}
U_{dm}=
-\frac12 G \int \int \frac{\sigma_{dm} (x) \sigma_{dm} (x') }{|x-x'|} dV dV'
\; .
\ee
The identity between Eqs. (\ref{lhs}) and (\ref{rhs}) is the
generalizaton of the
Layzer-Irvine equation \cite{[28]} describing how a collapsing
system reaches a state of dynamical equilibrium in
an expanding universe. One can see that the presence
of the coupling between dark energy and dark matter
changes both the time required by the system to reach
equilibrium, and the equilibrium configuration itself. For
a system in equilibrium ($\dot{K}_{dm} = \dot{U}_{dm} = 0$) we get the
condition:
\be
\label{LI}
(2-\bar\zeta) K_{dm} + (1+b_{em}) (1-2\bar\zeta) U_{dm} = 0 \; .
\ee
Taking $\bar\zeta = b_{em} = 0$ we recover the usual virial condition.

\section{MASS ESTIMATION AND LIMITS ON THE
COUPLING}

Galaxy clusters are the largest virialized structures
in the Universe, and their mass content is supposed to be
representative of the universe as a whole -- see, e.g., \cite{Ettori}.
They are composed of hundreds of galaxies, with the largest
fraction of their baryonic mass in the form of  hot, X-ray emitting
gas -- not stars.
Clusters are widely believed to be totally dominated by dark matter
\cite{Bahcall}, and are conspicuous: existing surveys have already
detected many thousands of clusters, and upcoming surveys will map
much more.

Cluster masses can be estimated in a variety of ways.
Weak lensing methods use the distortion in the pattern of
images behind the cluster (which acts as a lens)
to compute the projected gravitational
potential due to that cluster. Knowing the distances
to the cluster and to the background images, one can
derive the mass that causes
that potential. An independent mass estimation can be obtained from 
X-ray observations if we assume that the
ionized gas is in hydrostatic equilibrium. In this case 
the cluster mass can be determined by the
condition that the gravitational attraction is supported by the gas
pressure. Finally, we can measure radial velocities and the projected
distribution of cluster galaxies and, by
assuming that clusters are virialized,
one can infer their masses using the fact that 
$U \propto \sigma^2$ but $K \propto \sigma$, 
hence $U/K \propto M$.

However, Eq. [\ref{LI}] tells us that 
when there is coupling between dark matter
and dark energy, the equilibrium condition depends on
the coupling as:
\be
\label{equil}
(1+b_{em}) \frac{U_{dm}}{K_{dm}}
= -2 \frac{1-\bar\zeta/2}{1-2\bar\zeta} \; .
\ee
Hence, the mass that is estimated under the assumption that
$\bar\zeta=0$ is biased with the respect to the actual mass by a factor
of $ (1-\bar\zeta/2)/(1-2\bar\zeta)$.
 
One can compare directly the mass obtained through the virial theorem
with that determined by other methods.
Notice that the total mass of a cluster is, within our approximations, 
the integral of $\sigma_{Tot} = (1 + b_{em}) \sigma_{dm}$ over the
volume; hence,  the mass estimated through X-rays and weak lensing 
is related to the virial mass as $M = (1+b_{em}) M_{dm}$.

\begin{figure}[htb]
\vspace{0.5cm}
\includegraphics[width=8cm]{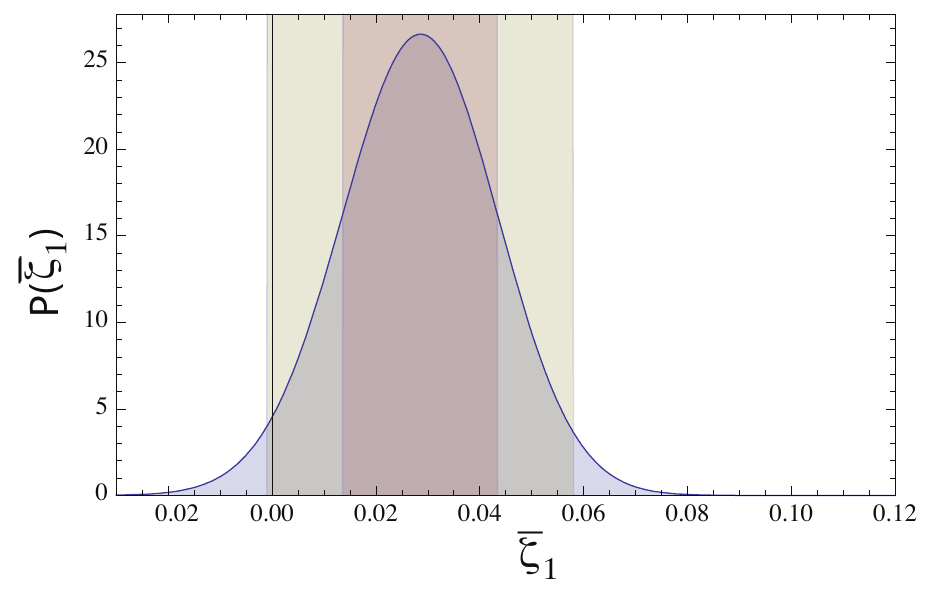}
\includegraphics[width=8cm]{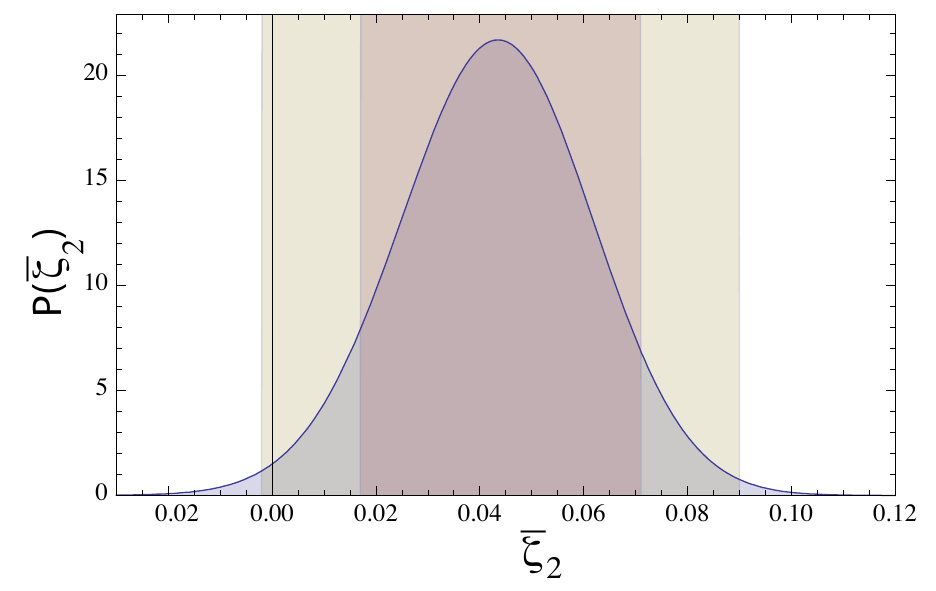}
\includegraphics[width=8cm]{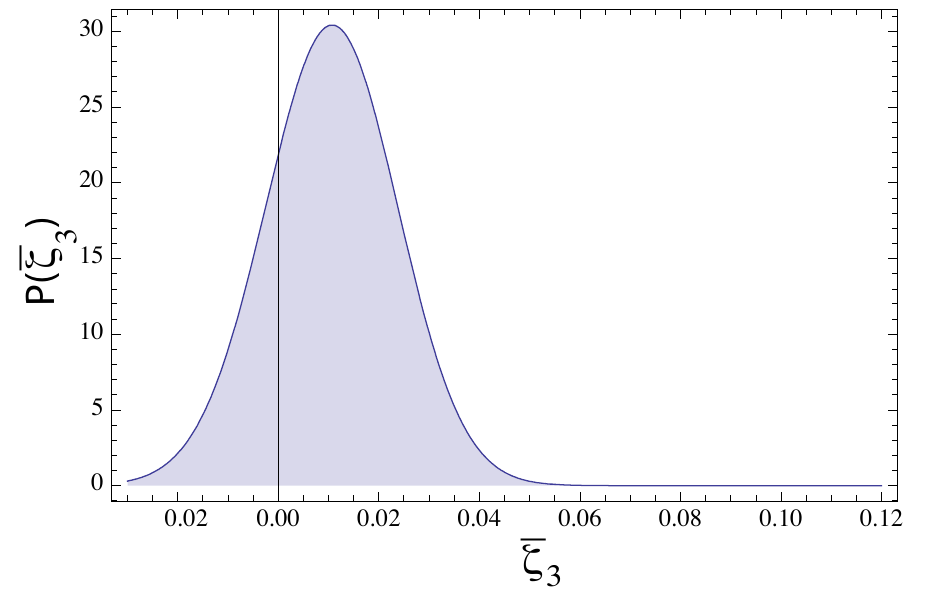}
\caption{Normalized probability distribution functions for 
the tests $f_1$ (top panel), $f_2$ (middle panel) and $f_3$ (bottom panel.) 
The shadowed regions on the top and middle panels indicate the 67\% and
95\% C.L. limits. }
\label{Fig1}
\end{figure}

Therefore, comparing the masses estimated through the {\bf conventional}
virial hypothesis with the masses estimated either by
weak lensing or by X-ray data, we get:
\be
\label{Mvir}
M_{X}  = M_{WL} =
M_{vir} 
\times \frac{1-\bar\zeta/2}{1-2\bar\zeta} 
\ee
Hence, there are three
tests one can make with these datasets:
\begin{eqnarray}
\label{f1}
f_1 &=& M_X/M_{vir} \; ,
\\
\label{f2}
f_2 &=& M_{WL}/M_{vir} \; ,
\\
\label{f3}
f_3 &=& M_{X}/M_{WL} \; .
\end{eqnarray}
If our interaction model is right, the 
first two tests, $f_1$ and $f_2$, should agree with each
other and put similar limits on the effective coupling parameter $\bar\zeta$,
while the third test, $f_3$, should only be a check of our
method, and its value should be equal to one 
unless there are unknown systematics affecting our mass estimates.
Notice that either 
a violation of the equivalence principle for dark matter or
a self-interaction of dark matter with itself, such
as suggested by \cite{KK06}, could also be tested by 
comparing the different mass estimates -- see also \cite{[27]}.

To what extent the data currently available on cluster masses allows us
to constrain the coupling parameter?  To compare masses obtained with the
different methods, we have assumed that the mass profile of the clusters 
is described by a Singular Isothermal Sphere (SIS). The main reasons are that
weak-lensing mass estimations need to adopt
a parametric model for the mass distribution in order to avoid the so-called
mass-sheet degeneracy \cite{Cypriano1}; this model is largely adopted in
weak lensing studies,  and most weak-lensing and X-ray mass 
estimations are possible only for radii significantly smaller than the virial.
The main advantage of this model is that it has a single parameter --
the velocity dispersion along the line-of sight $\sigma_v$ -- 
which can be easily determined: directly
from the observed radial velocities in the virial estimation; from
the X-ray temperature, [$\sigma_X^2 = kT_X/(\mu m_H)$,  where $\mu$=0.61 is
the mean molecular weight]; and from the fitting of the shear field in the
case of weak-lensing. Since in this model the mass inside
a given (projected) radius $R$ is $M(<R)=\pi \sigma^2 R/G$, 
to compare the masses obtained by each method we need only to compare the 
velocity dispersions. 

For this exercise, we
have analyzed data from galaxy clusters studied in
\cite{Cypriano1,Cypriano2,Hoekstra}.
Our sample has 33 clusters and was selected due to the homogeneity in 
the analysis procedure and avoiding clusters with 
evidence of dynamical activity, like substructures.
Ref. \cite{Cypriano1} presents a weak-lensing
analysis of 24 galaxy clusters also observed in X-rays,
verifying that clusters whith a hot intergalactic medium ($T_X > 8$ keV)
are very active. For our analysis, we selected from this paper 14 clusters
(from the 15 clusters with X-ray temperatures lower than 8 keV,
Abell 1651 also has evidence of significant substructure \cite{flin}).  
The cluster A586, discussed in \cite{Cypriano2}, also seems to be in 
equilibrium. The remaining 18 clusters of
our sample comes from \cite{Hoekstra}.

We have used this dataset to test 
the theory that the usual virial mass is biased by a factor
$(1-2\bar\zeta)/(1-\bar\zeta/2)$ when compared to other mass estimates. 
Although the three datasets have asymmetric errors, 
we have assumed that the likelihood function
associated with the three tests is symmetric, with width
$\sigma=\sqrt{\sigma_+\sigma_-}$. 
With these assumptions, the likelihood function of test $i$ is:
\be
\label{Like}
{\cal{L}}_i \propto
\prod_{n=1}^{N_i} \exp 
\left\{  - \frac{1}{2 \, \sigma_i^2 (n)} 
\left[
\frac{1-2\bar\zeta_i}{1-\bar\zeta_i/2} f_i(n)
\right]^2
\right\} \; ,
\ee
where the product runs over the data for each galaxy cluster $n$.
If our model is correct we should get 
$\bar\zeta_1 = \bar\zeta_2 =\bar\zeta $, 
and $\bar\zeta_3 = 0$.

In the three panels of Fig. 1 we show the probability distribution functions
for $f_1$, $f_2$ and $f_3$,
all computed for the top-hat prior $-0.2<\bar\zeta_i<0.2$,
which is more than sufficient to include the 3-$\sigma$ limits 
$-0.12<\bar\zeta_i<0.06 $ found by \cite{[19]}.
The shadowed regions in the top and middle panels mark the 67\% and 
95\% Confidence Level (C.L.) limits for $\bar\zeta_1$ and $\bar\zeta_2$.

For both the $f_1$ and $f_2$ tests 
we get a best-fit value of $\bar\zeta \sim 0.03-0.04$,
while for the check $f_3$ we indeed get that $\bar\zeta_3$ is consistent 
with zero with a high statistical significance.
Our 95\% C.L. limits are $0.0 \lesssim \bar\zeta \lesssim 0.06$ 
for the
test $f_1$ and $0.0 \lesssim \bar\zeta \lesssim 0.09$ for $f_2$.
This compares favourably with the 95\% C.L. constraint 
$-0.095 \lesssim \bar\zeta \lesssim 0.035$ obtained by \cite{[19]}.

For $f_1$ and $f_2$ we obtain that the null 
hypothesis ($\bar\zeta=0$) is marginally consistent
with the data, at the edge of the 95\% C.L. region. 
The statistical improvement
between the null hypothesis and the best-fit model with coupling 
is weak, though: we get a $\Delta \chi^2/{\rm d.o.f.} \sim 0.2$ 
for both the $f_1$ and $f_2$ tests.

We have also computed the Bayesian Information
Criterion (BIC) \cite{Liddle} to weigh if, and by how much, a coupling is
necessary. 
For the test $f_1$ we get $\Delta_{\rm BIC}^1 \approx -0.5$, while for the test
$f_2$ we get $\Delta_{\rm BIC}^2 \approx -2.1$. We can also estimate
the level of systematic uncertainties that would turn the BIC against our model
(i.e., when $\Delta_{\rm BIC}=0$):
an enhancement of 20\% of all uncertainties would make 
$\Delta_{\rm BIC}^1 \approx 0$, but still $\Delta_{\rm BIC}^2 \approx -1.2$; 
in order
to make $\Delta_{\rm BIC}^2 \approx 0$ it would take 
an enhancement of 70\% of the uncertainties.

The reliability of these constraints, however, are disputable, due to possible 
systematic effects in the mass determinations. For example, a virial mass 
estimate is affected by the assumptions about the galaxy orbits, cluster 
morphology, mass distribution, identification of interlopers, etc.
\cite{Hoekstra, Sodre, deFilippis}, 
and the robustness  of our simple SIS model does not mean 
that it is insensitive to (unknown) systematics.  
We can have a hint on the impact of these effects
for the constraints on the effective coupling constant by increasing its 
possible range of variation. For this exercise we have just redone the 
analysis assuming that the actual errors are twice the internal errors. 
In this case the original results ($\bar\zeta_{1} = 0.029 \pm 0.015$ and 
$\bar\zeta_{2} = 0.044 \pm 0.027$ at the 68\% C.L.) change to
$\bar\zeta_{1} = 0.029 \pm 0.030$ and 
$\bar\zeta_{2} = 0.044 \pm 0.037$, i.e.,
the most probable value of $\bar\zeta_{1,2}$ does not
change (as expected in this case), but the error in the estimates almost
doubles, reducing to about one sigma the level of detection of a non-zero
coupling constant. Moreover, since we can only constrain the effective
coupling parameter $\bar\zeta = (1+b_{em}) \zeta$, the bias between
dark matter and dark energy in virialized structures
could enhance (if $b_{em} > 0$) or suppress ($b_{em} <0$) 
our ability to constrain the true coupling $\zeta$.

These results show that the reliability of a detection
of the effective coupling parameter requires very good knowledge of possible 
systematic errors. Nevertheless, it also shows that
if in the future we can produce a sample with reliable mass estimates and
controlled systematics, we will indeed be able to constrain $\bar\zeta_{1,2}$
and verify whether this hint of a coupling between dark matter and dark
energy found with current data is confirmed.

\section{CONCLUSIONS}

We have estimated the effective coupling between dark energy and dark matter
through the internal dynamics of galaxy clusters. In the presence of
coupling, the flow of mass and energy between the components changes
the virial condition in a way that can be tested by comparing
different estimators for the mass of clusters. We searched for this
signature in 33 galaxy clusters for which 
reliable X-ray, weak lensing and optical data were available.

Our results indicate a weak preference for a small but positive 
effective coupling
constant $\bar\zeta$ -- in line with predictions made by some of us
\cite{[12],[13]}. Since the statistical significance is still 
low ($\Delta \chi^2/{\rm d.o.f.} \sim 0.2$), it is paramount that more
clusters (with homogeneous mass determinations and good control of 
systematics) be tested. If a significant indication of such coupling is 
still found, this
would open a tantalizing new window on the nature of the dark sector.

\section*{Acknowledgments}
This work has been supported by FAPESP and CNPq
of Brazil, by NNSF of China, Shanghai Education
Commission, and Shanghai Science and Technology Commission.



\begin{thebibliography}{99}



\bibitem{cosmoaccel} A. G. Riess et al., {\it Astron. J.} {\bf 116}: 1009 (1998);
  S. Perlmutter et al., {\it Astrophys. J.} {\bf 517}: 565 (1999); 
  P. de Bernardis et al., {\it Nature} {\bf 404}: 955 (2000); 
  R. Knop et al., {\it Astrophys. J.} {\bf 598}: 102 (2003); 
  A. G. Riess et al., {\it Astrophys. J.} {\bf 607}: 665 (2004); 
  P. Astier et al., {\it Astron. Astrophys.} {\bf 447}: 31 (2006); 
  W. M. Wood-Vasey et al., {\it Astrophys. J.} {\bf 666}: 694 (2007), astro-ph/0701041.

\bibitem{wmap} WMAP, D. N. Spergel et al., {\it Astrophys. J. Suppl.} {\bf 170}:
377 (2007).

\bibitem{bao} SDSS, D. J. Eisenstein et al., {\it Astrophys. J.} {\bf 633}: 560
(2005).

\bibitem{darkenergy} T. Padmanabhan, {\it Phys. Rept.} {\bf 380}: 235 (2003), 
hepth/0212290; P. J. E.Peebles and B. Ratra, {\it Rev. Mod. Phys.}
{\bf 75}: 559 (2003), astro-ph/0207347; 
V. Sahni, {\it Lect. Notes Phys.} {\bf 653}: 141 (2004), astro-ph/0403324.

\bibitem{amendola} L. Amendola, {\it Phys. Rev.} {\bf D62}: 043511 (2000); 
Luca Amendola and Claudia Quercellini, {\it Phys. Rev.} {\bf D68}: 023514 (2003); 
Luca Amendola, Shinji Tsujikawa and M. Sami, {\it Phys. Lett.} {\bf B632}: 155 (2006).

\bibitem{[6]} M. Pietroni, {\it Phys. Rev.} {\bf D67}: 103523 (2003); 
D. Comelli, M. Pietroni and A. Riotto, {\it Phys. Lett.} {\bf B571}: 115 (2003).

\bibitem{[7]} G. Farrar and P. J. E. Peebles, {\it Astrophys. J.} {\bf 604}: 1 (2004).

\bibitem{[8]} D. Pavon, W. Zimdahl, {\it Phys. Lett.} {\bf B628}: 206 (2005), 
gr-qc/0505020.


\bibitem{[10]} S. Campo, R. Herrera, G. Olivares and D. Pavon, 
{\it Phys. Rev.} {\bf D74}: 023501 (2006);
S. Campo, R. Herrera and D. Pavon, {\it Phys. Rev.}  {\bf D71}: 123529 (2005); 
G. Olivares, F. Atrio- Barandela and D. Pavon, {\it Phys. Rev.} {\bf D71}: 063523 (2005).

\bibitem{[11]} G. Olivares, F. Atrio-Barandela, D. Pavon, {\it Phys. Rev.}
{\bf D74}: 043521 (2006).


\bibitem{[12]} B. Wang, Y. G. Gong and E. Abdalla, {\it Phys. Lett.} {\bf B624}: 141 (2005); B. Wang, C. Y. Lin and E. Abdalla, {\it Phys. Lett.} {\bf B637}: 357 (2006).

\bibitem{[13]} B. Wang, J. Zang, C.Y. Lin, E. Abdalla and
S. Micheletti, {\it Nucl. Phys.} {\bf B778}: 69 (2007).

\bibitem{[14]} S. Das, P. S. Corasaniti and J. Khoury,
{\it Phys. Rev.} {\bf D73}: 083509 (2006).

\bibitem{[15]} G. Huey and B. Wandelt, {\it Phys. Rev.} {\bf D74}: 023519
(2006).

\bibitem{[16]} W. Zimdahl, {\it Int. J. Mod. Phys.} {\bf D14}: 2319 (2005),
gr-qc/0505056.

\bibitem{[17]} E. Abdalla and B. Wang {\it Phys. Lett.} {\bf B651}: 89
(2007).

\bibitem{[18]} C. Feng, B. Wang, Yungui Gong and Ru-Keng Su,
{\it JCAP} {\bf 0709}: 005 (2007), arXiv:0706.4033.

\bibitem{[19]} Z. K. Guo, N. Ohta and S. Tsujikawa, 
{\it Phys. Rev.} {\bf D76}: 023508 (2007), astro-ph/0702015.

\bibitem{[20]} P. Ferreira and M. Joyce, {\it Phys. Rev.} {\bf D58}: 023503
(1998).

\bibitem{[21]} P. Viana and A. Liddle, {\it Phys. Rev.} {\bf D57}: 674 (1998).

\bibitem{[22]} L. R. Abramo and F. Finelli, {\it Phys. Rev.} {\bf D64}: 083513
(2001).

\bibitem{[23]} R. Bean, {\it Phys. Rev.} {\bf D64}: 123516 (2001).

\bibitem{[24]} N. Nunes and D. Mota, {\it Mon. Not. Roy. Astron. Soc.} {\bf 368}:
751 (2006).

\bibitem{[25]} L. R. Abramo, R. C. Batista, L. Liberato and R. Rosenfeld,
{\it JCAP} {\bf 0711}: 012 (2007), arXiv: 0707.2882 [astro-ph]

\bibitem{[26]} M. Manera and D. Mota, {\it Mon. Not. Roy. Astron. Soc.}
{\bf 371}: 1373 (2006).

\bibitem{[27]} O. Bertolami, F. Gil Pedro and M. Le Delliou
{\it Phys.Lett.} {\bf B654}: 165 ,(2007),astro-ph/0703462;
idem, arXiv:0705.3118 [astro-ph]; idem arXiv:0801.0201 [astro-ph]

\bibitem{[28]} P. J. E. Peebles, ``Physical Cosmology'' (Princeton U.
Press, Princeton, 1993).


\bibitem{Ettori} S. Ettori, {\it Mon. Not. Roy. Astron. Soc.}
{\bf 344}: L13 (2003).

\bibitem{Bahcall} N. Bahcall, in ``Formation of Structure in the Universe'', 
ed. A. Dekel \& J. P. Ostriker (Cambridge University Press, 1999).

\bibitem{KK06} M. Kesden and M. Kamionkowski,
{\it Phys. Rev. Lett.} {\bf 97}: 131303 (2006); 
idem, {\it Phys. Rev.} {\bf D 74}: 083007 (2006).

\bibitem{Cypriano1} E. S. Cypriano, L. Sodr\'e, J.-P. Kneib and L. E.
Campusano, {\it Astrophys. J.} {\bf 613}: 95 (2004).

\bibitem{Cypriano2} E. S. Cypriano,
G. B. Lima Neto, L. Sodr\'e, J.-P. Kneib and L. E. Campusano,
{\it Astrophys. J.} {\bf 630}: 38 (2005).

\bibitem{Hoekstra} H. Hoekstra, {\it Mon. Not. Roy. Astron. Soc.}
{\bf 379}: 317 (2007).


\bibitem{flin} P. Flin and J. Krywult, {\it Astron. Astrophys.}
{\bf 450}: 9 (2006).


\bibitem{Liddle} A. Liddle, {\it Mon. Not. Roy. Astron. Soc. Lett.} {\bf 377}:
L74 (2007).

\bibitem{Sodre} L. Sodr\'e, H. V. Capelato, J. E. Steiner, A. Mazure,
{\it Astronom. J.} {\bf 97}: 1279 (1989).

\bibitem{deFilippis} E. De Filippis, M. Sereno, M. W. Bautz, G. Longo,
{\it Astrophys. J.} {\bf 625}: 108 (2005).


\end{thebibliography}
\end{document}